\documentstyle[12pt,moriond,psfig]{article}

\def\ltsim{\raise0.3ex\hbox{\small$\lsim$}\,\,}
\def\gtsim{\raise0.3ex\hbox{\small$\gsim$}\,\,}

\begin{document}

\heading{BUILDING GALAXIES: CONFERENCE SUMMARY}

\vspace{6mm}
\begin{center}
S.~Michael Fall 

{\it Space Telescope Science Institute, Baltimore, USA}
\end{center}
\vspace{2mm}

\begin{moriondabstract}

This article summarizes---from a personal perspective---some of the main 
themes that have emerged at this conference and in this field generally in
the past few years. 

\end{moriondabstract}

\section{Introduction}

This field is now in an explosive phase of growth, driven mainly by a 
wealth of observations at ever higher redshifts. We probably know as 
much today about galaxies at $z=3$ as we did five years ago about galaxies 
at $z=0.3$. This new knowledge is the product of many clever people and some 
powerful telescopes and instruments: HST, Keck, COBE, ISO, and SCUBA among 
them. With these observations and some theoretical background, we have
begun to write an outline of the story of galaxy formation, the subject
of this conference. If we continue to make progress at anything like the 
present rate, we will know, within a decade or two, the full story of 
galaxy formation. At that time, I predict, we will look back to this one
and recall with nostalgia how exciting it was to help write that story.

We have heard at this meeting (and can now read in this book) a large 
number of excellent presentations, all with new data and/or ideas. While
this has certainly made the conference stimulating, it has not made the 
job of summarizing it easy! In fact, so many new results were presented 
that it would be impossible for me to review more than a small fraction 
of them. Instead, it is probably more valuable and tractable to highlight 
a couple of the major themes that pervade much of the recent work in this 
field and at this meeting. These are the global evolution of galaxies and 
the origin of the Hubble sequence; the first ignores the individuality of 
galaxies, while the second attempts to understand it. 

\section{Global Evolution of Galaxies}

One of the grand themes to emerge in the last few years is the idea 
that we may be able to trace the global histories of star formation, gas 
consumption, and metal production in galaxies from high redshifts to the 
present. By ``global,'' we mean averages over the whole population of 
galaxies or, equivalently, over large, representative volumes of the 
universe. This idea is illustrated in Figure~1, where we sketch the 
evolution of the contents of a large, comoving box. We can conveniently 
quantify the masses of the different constituents of the box by the 
corresponding mean comoving densities normalized to the critical density. 
We are especially interested in the comoving densities of stars, gas 
(both inside and outside galaxies, i.e., ISM and IGM), metals, dust, and 
black holes: $\Omega_{\rm_S}$, $\Omega_{\rm ISM}$, $\Omega_{\rm IGM}$, 
$\Omega_{\rm M}$, $\Omega_{\rm D}$, and $\Omega_{\rm BH}$, respectively. 
We are also interested in the cosmic emissivity $E_\nu$, the power 
radiated per unit comoving volume per unit rest-frame frequency $\nu$, 
and the background intensity $J_\nu$, the power received per unit solid 
angle of sky per unit area of detector per unit observed frequency $\nu$.

\begin{figure}[t]
\centerline{\psfig{file=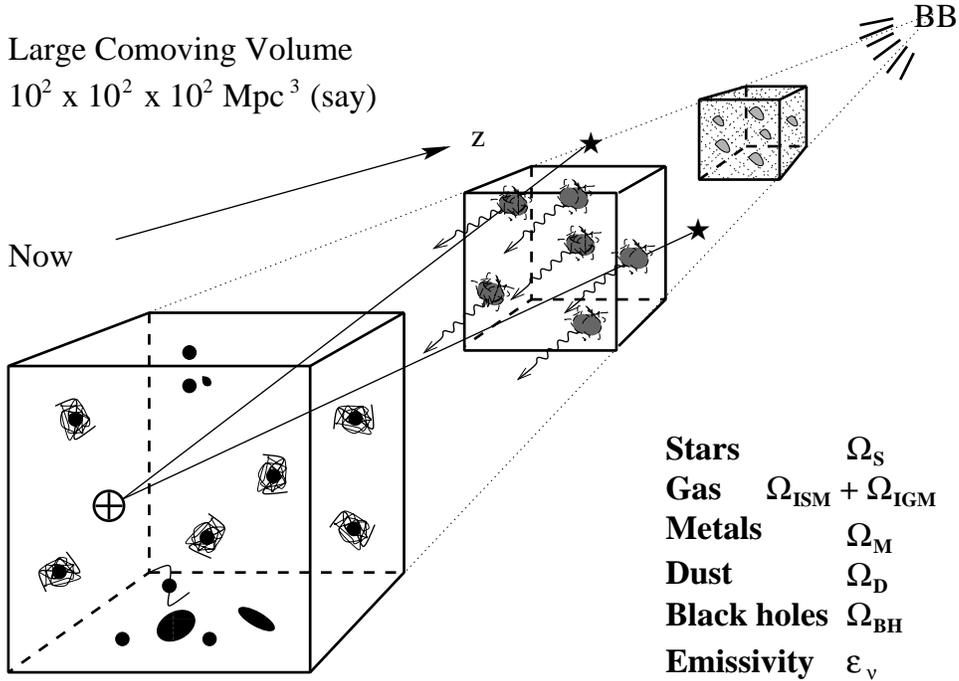,angle=270,width=5in}}
\caption{Evolution of the contents of a large comoving volume of the 
universe, from the big bang to the present, including galaxies and the
IGM. The wavy lines represent the light emitted by stars, AGN, and dust 
in galaxies; the straight lines represent the light emitted by quasars 
and then partially absorbed or scattered in the ISM of foreground 
galaxies and the IGM.}
\end{figure}

After recombination, our comoving box is filled with neutral, metal-free 
gas with nearly uniform density. Perturbations in this intergalactic medium 
(IGM) eventually condense, probably by gravitational clumping and inflow, 
into protogalaxies. Stars then form in the resulting interstellar medium 
(ISM). They produce metals and may drive outflows of gas from galaxies. In 
this way, both the ISM and IGM may be enriched with metals. Some of the 
metals remain in the gas phase; others condense into solid dust grains. 
Black holes form in the nuclei of some, perhaps even all, galaxies and, 
when fueled, can power active galactic nuclei (AGN). Some of the radiation
emitted by stars and AGN propagates freely, while the rest is absorbed and 
then emitted at longer wavelengths by dust. Thus, the radiation we detect 
from galaxies tells us primarily about their star, AGN, and dust contents.
The spectra of high-redshift quasars contain signatures of the absorption 
and scattering of radiation by the intervening ISM and IGM (absorption lines, 
reddening, etc). Such observations tell us primarily about the composition
and comoving densities of the ISM and IGM.

Exactly how all this happens is not yet known, of course. This is a
major long-term goal, the holy grail, of our subject. It should be clear 
from Figure~1 and the commentary above, however, that the constituents of 
our comoving box, including the radiation that propagates through it, are 
very much interrelated. In fact, the corresponding comoving densities must 
obey a series of coupled conservation equations. This is illustrated in 
Figure~2, which shows the hypothetical but plausible evolution of several 
of these comoving densities. Clearly, $\Omega_{\rm S}$, $\Omega_{\rm ISM}$, 
and $\Omega_{\rm IGM}$ must add up to $\Omega_{\rm baryon}$, a constant. 
Similarly, $\Omega_{\rm M}^{\rm S}$, $\Omega_{\rm M}^{\rm ISM}$, and 
$\Omega_{\rm M}^{\rm IGM}$, the comoving densities of metals in stars, the 
ISM and the IGM, must add up to $\Omega_{\rm M}$. Moreover, $\Omega_{\rm M}$ 
remains a fixed fraction of $\Omega_{\rm S}$ on the assumption that the 
global yield is constant and that delayed recycling is negligible (a 
good approximation in the present context). The bolometric emissivity, 
$E_{\rm bol}=\int E_\nu d\nu$, is the sum of two terms, one nearly 
proportional to the star formation rate $\dot\Omega_{\rm S}$ and one 
proportional to the black hole fueling rate $\dot\Omega_{\rm BH}$. 
The spectral shape of $E_\nu$ depends on $\dot\Omega_{\rm S}$, 
$\dot\Omega_{\rm BH}$, and the amount of reprocessing by dust and 
hence on $\Omega_{\rm D}$. Finally, the background and emissivity are 
related by $J_\nu=(c/4\pi)\int E_{(1+z)\nu} dt$.

\begin{figure}[t]
\centerline{\psfig{file=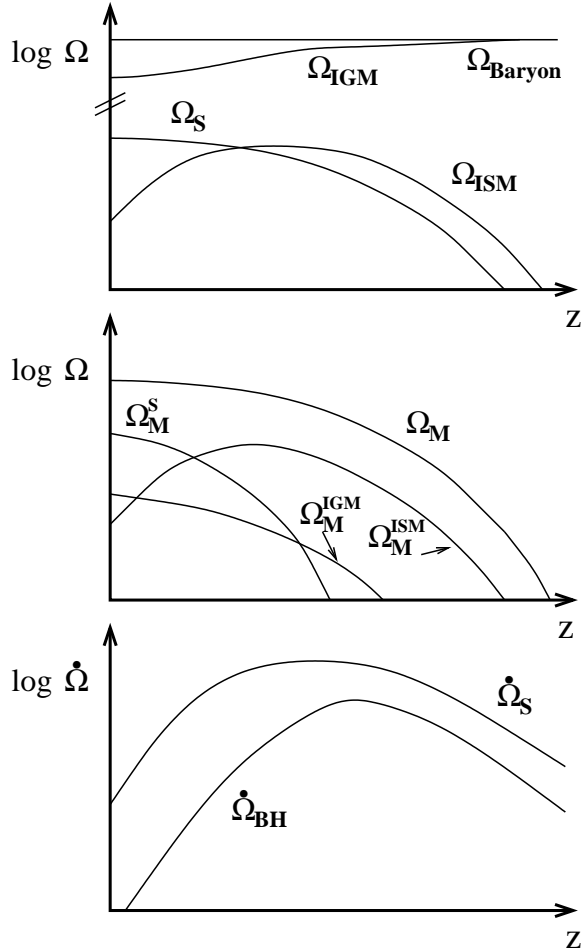,width=3in}}
\caption{Hypothetical evolution of the comoving densities of different 
constituents of the universe (as functions of redshift). Top: baryons, 
stars, ISM, and IGM. Middle: metals in total, in stars, in the ISM, and
in the IGM. Bottom: rates of star formation and black hole fueling.}
\end{figure}

In the past few years, we have made great progress in sketching out a
global picture of galactic evolution. In fact, we now have empirical 
estimates of several of the quantities mentioned above at redshifts from 
$z=0$ up to $z\approx4$ and hence over most of cosmic time. One of the 
great advantages of the global approach is that the quantities of interest 
are so interrelated that, in principle, any one of them could be predicted 
from the others. For example, one could infer the global history of star 
formation from its consequences on the ISM and IGM, and hence on the 
spectra of background quasars, without observing a single photon emitted 
by a star! In practice, of course, there are uncertainties. Some of these 
stem from the difficulty of making measurements at particular wavelengths 
and redshifts. For example, it is harder to determine the redshifts of 
galaxies at $1\ltsim z\ltsim 2$ than at $z\ltsim1$ and $z\gtsim2$. 
Another source of uncertainty is that various differential quantities 
must be extrapolated outside the ranges over which they have been observed 
in order to estimate integral quantities. The emissivity and background, 
for example, are computed from the luminosity function and number vs flux 
density relation through $E_\nu = \int L_\nu \phi(L_\nu) dL_\nu$ and 
$J_\nu = \int S_\nu (dN/dS_\nu) dS_\nu$. As Figure~3 indicates, 
extrapolating $\phi(L_\nu)$ to $L_\nu \rightarrow 0$ and $dN/dS_\nu$ to 
$S_\nu \rightarrow 0$ may introduce uncertainties in $E_\nu$ and $J_\nu$.

\begin{figure}[t]
\centerline{\psfig{file=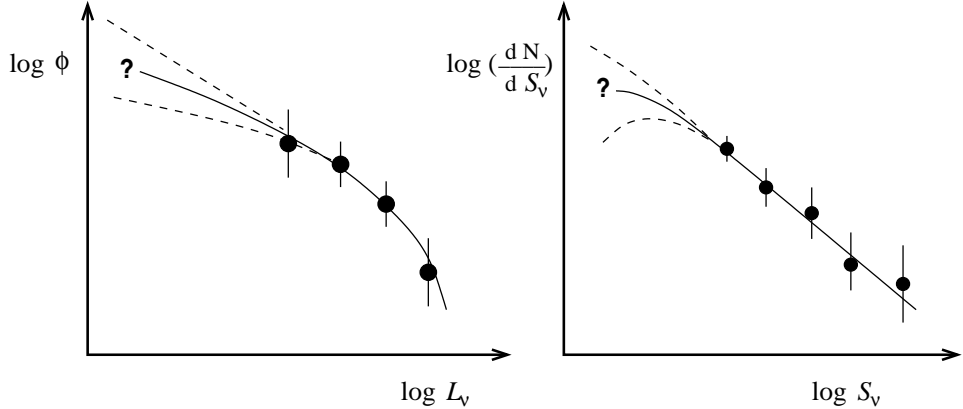,angle=270,width=5in}}
\caption{Illustrative extrapolations at the faint ends of the luminosity 
function and number vs flux density relation. These cause uncertainty in 
estimates of the cosmic emissivity $E_\nu$ and background intensity $J_\nu$.}
\end{figure}

How well are we doing with this program? Most authors seem to agree that 
the global star formation rate $\dot\Omega_{\rm S}$ declines by a large 
factor ($\sim\! 10$) from $z\approx1$ to $z=0$, although there are still 
some uncertainties (i.e., factors of 5--20 are possible). It appears that 
$\dot\Omega_{\rm S}$ may level off at $1 \ltsim z \ltsim 2$, but its behavior 
at higher redshifts has been a topic of much recent debate. Within the 
large uncertainties, it could rise, fall, or remain constant. Part of the 
uncertainty comes from the unknown corrections for dust when
converting observed UV emissivities into star formation rates. Some 
authors have ignored these corrections altogether, while others have
advocated corrections by an order of magnitude or more. In this context, 
it is worth noting that the average correction factor (over all redshifts) 
can be estimated by comparing the energy densities in the background at 
wavelengths above and below 10\,$\mu$m:
$$
CF\approx 1 + \bigg( \int_{\lambda>10\mu{\rm m}} J_\nu d\nu \bigg/ 
\int_{\lambda<10\mu{\rm m}} J_\nu d\nu \bigg).
$$
Figure~4 shows recent estimates of, and limits on, $J_\nu$ over a wide 
range of wavelengths. From this, we infer a modest correction factor,
$CF\approx$ 2--3 (i.e., neither negligible nor dominant).

\begin{figure}[t]
\centerline{\psfig{file=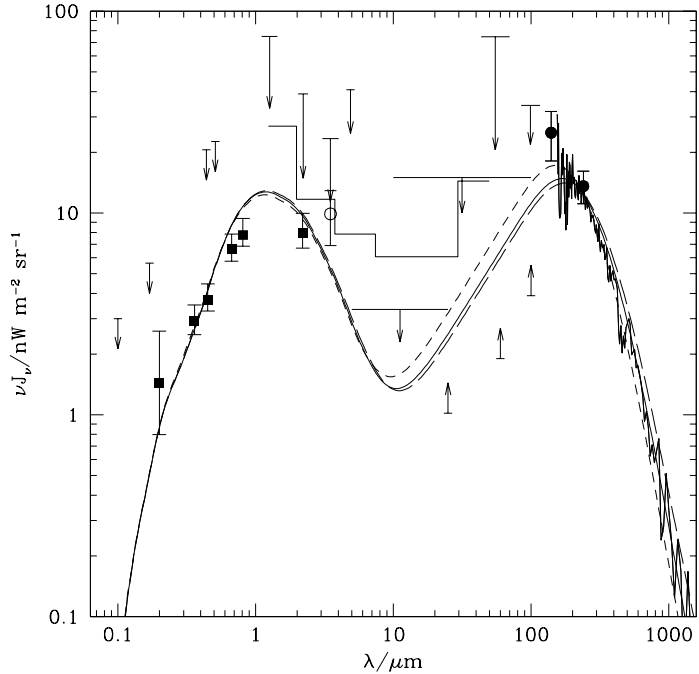,width=4in}}
\caption{Extragalactic background intensity times frequency plotted
against wavelength. The filled squares are from galaxy counts, while the 
filled circles and zigzag line are the DIRBE and FIRAS measurements, 
respectively. The arrows and stepped lines indicate various 
observational limits. The curves are from models of cosmic chemical
evolution by Pei, Fall, \& Hauser.}
\end{figure}

Damped Ly$\alpha$ absorbers (DLAs) are usually taken to represent the ISM 
of galaxies. The reasons for this are that they contain most of the neutral
gas in the universe and have column densities near or above the threshold
for star formation ($N_{\rm H} \gtsim 10^{20}$\,cm$^{-2}$). The comoving 
density of gas in the DLAs declines by a factor of about 10 between $z=$ 
2--3 and $z=0$, the redshifts at which it is known most reliably. Moreover, 
$\Omega_{\rm ISM}$ at $z=$ 2--3 is nearly equal to $\Omega_{\rm S}$ at 
$z=0$, highly suggestive of the conversion of ISM into stars. The 
value of $\Omega_{\rm ISM}$ at $z\sim1$ is less certain because relatively
few DLAs are known at these redshifts. The global mean ISM metallicity, 
$Z_{\rm ISM}\equiv\Omega_{\rm M}^{\rm ISM}/\Omega_{\rm ISM}$, rises 
substantially, from $Z_{\rm ISM}\approx0.1\,Z_\odot$ at $z=$ 2--3 to 
$Z_{\rm ISM}\approx Z_\odot$ at $z=0$, again suggestive of a great deal of 
star formation in this period. However, the observed value of $Z_{\rm ISM}$ 
at $z\sim1$ is surprisingly low. This may be the result of a selection 
effect in which metal-rich regions of the ISM are also dust-rich and 
hence obscure any quasars behind them. Finally, we note recent estimates
of the global mean IGM metallicity: $Z_{\rm IGM}\equiv
\Omega_{\rm M}^{\rm IGM}/\Omega_{\rm IGM}\sim10^{-3}\,Z_\odot$ at 
$z\approx3$. This and the observations of DLAs indicate that the comoving
density of metals in the IGM is less than that in the ISM at this redshift. 
We know very little about the metallicity of the IGM at $z \ltsim 1$,
a situation we hope will improve soon. 

Another recent development is the realization that AGN may make a
significant contribution to the global radiation budget. This can be 
understood as follows. The present comoving density of black holes 
can be estimated from dynamical studies of the nuclei of nearby galaxies.
Given an efficiency of conversion between black hole mass and radiant energy 
and a typical redshift of conversion, one can estimate the contribution of 
AGN to the bolometric background intensity. For $\epsilon\sim10$\% and
$z\approx2$, both plausible values, the result is $J_{\rm AGN,bol}\sim0.2
J_{\rm bol}$. Within the uncertainties in the various input quantities
(including $\Omega_{\rm BH}$ at $z=0$), $J_{\rm AGN, bol}$ could be
several times larger or smaller, i.e., nearly dominant or nearly 
negligible. For comparison, AGN make only a minor contribution to 
$J_\nu$ at visible wavelengths. Thus, if they are a major contributor
to $J_{\rm bol}$, most of the radiation must be reprocessed by dust. 
Visible and radio observations of the sub-mm sources detected with SCUBA 
may help to resolve this issue. In any case, it serves as a useful reminder 
that the observed background is strictly an upper limit on the mean 
intensity of stellar radiation.

\section{Origin of the Hubble Sequence}

The other major theme on which we have seen many new results is 
the origin and evolution of the Hubble sequence of galactic morphologies.
Much of this progress comes from deep imaging with HST at visible and,
very recently, near-infrared wavelengths. 
From these and other observations, the following picture emerges.
Large elliptical galaxies in clusters appear relatively old at 
$z\approx1$, indicating formation at $z \gtsim $ 2--3.
In the field, however, a small but significant fraction of elliptical 
galaxies and the spheroids of disk galaxies ($\sim\! 20$\%) either formed 
late or experienced recent episodes of star formation (at $z \ltsim 1$).
Large galactic disks appear to have changed relatively little---in 
luminosity, size, or rotation velocity---from $z\approx1$ to $z=0$.
The decline in the cosmic UV emissivity between $z\approx1$ and $z=0$ 
is caused mainly by rapid evolution in the number density and/or 
luminosities of small galaxies [e.g., blue compact galaxies (BCGs)].
The last result needs confirmation, however, because it depends on the
faint end of the luminosity function, which is notoriously difficult 
to determine.

The situation at higher redshifts is potentially even more interesting: 
galaxies at $z \gtsim 2$ appear smaller and more disturbed than their
present-day descendants.
We observe some normal-looking elliptical galaxies but remarkably few 
(if any) normal-looking disk galaxies. 
Taken at face value, these observations suggest that most ellipticals and 
spheroids formed at $z \gtsim $ 2--3, that most disks (large ones, at least) 
formed later, at $1 \ltsim z \ltsim 2$, that the subsequent evolution of
large galaxies was mainly passive (ellipticals and spheroids) or quiescent
(disks), while the activity of dwarf galaxies declined at $z \ltsim 1$.
If this picture is correct, the Hubble sequence, in the form familiar to 
us, largely came into being during the period $1 \ltsim z \ltsim 2$.
It is highly significant, in this regard, that the sizes and morphologies 
of galaxies at high redshifts appear much the same at visible and 
NIR wavelengths (i.e., rest-frame UV and visible wavelengths). 
There is, however, a selection effect to worry about.
As a result of the usual cosmological dimming, it becomes increasingly
difficult to observe low surface brightness features, such as the outer 
parts of quiescent disks, at higher redshifts.
We need to understand this effect better before we can be confident we 
have witnessed the origin of the Hubble sequence. 

Tentative though these results may be, they do invite some harmless
speculation. 
First, the observations are at least qualitatively consistent with the 
idea that galaxies formed in a hierarchical sequence, starting with small
objects and progressing to larger ones by merging and inflow.
Second, the observations contain a vital clue as to why galactic disks 
appeared relatively late.
At high redshifts, galaxies were close together and interacted frequently, 
leading in some cases to mergers and the formation of elliptical galaxies 
and spheroids (``hot'' stellar systems).
These disturbances, clearly visible in the HST images, would destroy or 
prevent the formation of thin disks (``cold'' stellar systems).
At lower redshifts, however, galaxies were farther apart and interacted
rarely, permitting the formation and survival of disks. 
In this situation, the gas in galactic halos can cool, contract, spin up, 
and settle into thin, rotationally supported disks, where it can then
be converted into stars. 
Clearly, much remains to be done to confirm or refute this picture.
But we can look forward to a bright future, informed by even better
observations with HST, with the large ground-based telescopes, and 
ultimately, with NGST.

\section{Appreciation}

This has been a wonderful conference: excellent food, excellent skiing, 
excellent conversation, and especially, excellent science. The success
of the meeting derives from the efforts of many people, including all
the speakers and participants. We are especially grateful to the 
scientific organizing committee---V\'eronique Cayatte, Bruno Guiderdoni, 
Fran\c{c}ois Hammer, Trinh Xuan Thuan---and the organizer of 
organizers---Tr\^an Thanh V\^an---for the vision that created this and 
other Moriond meetings. Last but not least, we thank Sabine Kimmel, who 
helped turn this vision into reality.

\vfill
\end{document}